\documentstyle[TSF11-ed]{article}
\newcommand{\Avr}[1]{\langle{#1}\rangle}

\newcommand{\Vec}[1]{\mbox{\boldmath ${#1}$}}
\begin{document}
\small
\title{\sffamily INVESTIGATION OF RENORMALIZATION GROUP METHODS FOR THE
NUMERICAL SIMULATION OF ISOTROPIC TURBULENCE}

\author{
\vv{\bfseries\sffamily
David McComb, Taek-Jin Yang, Alistair Young}\\
\vv
Department of Physics and Astronomy\\
\vv
University of Edinburgh\\
\vv
James Clerk Maxwell Building\\
\vv
Mayfield Road\\
\vv
Edinburgh EH9 3JZ\\
\vv
United Kingdom\\\\
\vv{\bfseries\sffamily
Luc Machiels}\\
\vv
Laboratory of Fluid Mechanics\\
\vv
Swiss Federal Institute of Technology\\
\vv
CH-1015 Lausanne\\
\vv
Switzerland}

\maketitle

\section{INTRODUCTION}

Over the years, our research into turbulence at Edinburgh has
concentrated on the application of renormalization methods to the
prediction of the energy spectrum of isotropic turbulence. General
discussions of this work will be found elsewhere (McComb 1990, 1995),
while accounts of specific progress have been given previously in this
conference series (McComb \& Shanmugasundaram 1983, McComb, Filipiak,
Roberts \& Watt, 1991).

From a practical point of view, the most promising development in this
area is undoubtedly {\bf Renormalization Group} or {\bf RG}. If we work
in the Fourier representation, in principle, this involves the
progressive averaging out of high-wavenumber modes in bands, with
rescaling at each step, until a fixed point is reached. The result is,
in effect, a `subgrid model' for large-eddy simulation. 

RG has enjoyed its successes in other areas of statistical physics.
However, its application to turbulence faces several technical
difficulties, which have to be circumvented by uncontrolled
approximations. Indeed, in view of the deterministic nature of the
Navier-Stokes equations, it is clear that the operation of averaging out
the high-wavenumber modes while keeping the low-wavenumber modes
constant, cannot be done rigorously and in itself can only be an
approximation.

With points like this in mind, we have recently adopted direct numerical
simulation as a tool for probing the basic feasibility of using RG
techniques to reduce the number of degrees of freedom requiring to be
numerically simulated. In this paper, we present some of the first
results of this approach. We begin by discussing the RG approach in detail. 

\section{RENORMALIZATION GROUP THEORY}

\subsection{Basic Equations}

Working in Fourier-wavevector ($\bf k$) space and restricting our
attention to turbulent velocity fields which are homogeneous, isotropic
and stationary, we may write the pair-correlation of velocities as
\begin{equation}
\langle u_{\alpha}({\bf k},t) u_{\alpha}({\bf k}', t') \rangle
  = Q(k,t-t')D_{\alpha\beta}({\bf k})\delta({\bf k}-{\bf k}'),
\end{equation}
where $Q(k,t-t')$ is the spectral density and the projector
$D_{\alpha\beta}({\bf k})
=\delta_{\alpha\beta} + k_{\alpha}k_{\beta}k^{-2}$ arises due to
the incompressibility condition. Thus, the energy spectrum
$E(k)=4\pi k^{2} Q(k)$ with $Q(k)=Q(k,0)$ and the maximum cut-off
wave-number, $k_0$, is defined via the dissipation integral
\begin{equation}
\varepsilon
  =      \int_{0}^{\infty}dk~2\nu_0 k^2 E(k)
  \simeq \int_{0}^{k_0}dk~2\nu_0 k^2 E(k),
\end{equation}
where $\varepsilon$ is the dissipation rate, $\nu_0$ is the kinematic
viscosity, and $k_0$ is of the same order of magnitude as the Kolmogorov
dissipation wave-number.

\subsection{Renormalization Group Theory}

Taking our goal to be the calculation of the energy spectrum $E(k)$,
our intermediate objective is to find an analytical method of
reducing the number of degrees of freedom (or Fourier modes), in order
to make the numerical solution of the equations of motion a practical
proposition. Let us consider how this might be done by using  RG.

First, we divide up the velocity field at $k=k_1$ as
$u_{\alpha}({\bf k},t)=u^{-}_{\alpha}({\bf k},t)$ for $0<k<k_1$ and
$u_{\alpha}({\bf k},t)=u^{+}_{\alpha}({\bf k},t)$ for $k_1<k<k_0$,
where $k_1=(1-\eta)k_0$ and the bandwidth parameter $\eta$ satisfies
the condition $0<\eta<1$. Working with the standard form of the
solenoidal Navier-Stokes equation in $k$-space, we may write the
evolution of the low-$k$ velocity field for $0<k<k_1$ as
\begin{eqnarray}
\lefteqn{\left[\frac{\partial}{\partial t}+\nu_{0} k^2\right]
         u^{-}_{\alpha}({\bf k},t)
        }
  \nonumber \\
  &=& M^{-}_{\alpha\beta\gamma}({\bf k})\int d^3 j~
      \Big[
      u^{-}_{\beta}({\bf j},t)u^{-}_{\gamma}({\bf k}-{\bf j},t)
  \nonumber \\
  &+& 2u^{-}_{\beta}({\bf j},t)u^{+}_{\gamma}({\bf k}-{\bf j},t)
      + u^{+}_{\beta}({\bf j},t)u^{+}_{\gamma}({\bf k}-{\bf j},t)
      \Big],
\label{NSE:um}
\end{eqnarray}
and the evolution of the high-$k$ velocity field for the first shell,
$k_1<k<k_0$, as
\begin{eqnarray}
\lefteqn{\left[\frac{\partial}{\partial t}+\nu_{0} k^2\right]
         u^{+}_{\alpha}({\bf k},t)
        }
  \nonumber \\
  &=& M^{+}_{\alpha\beta\gamma}({\bf k})\int d^3 j~
      \Big[
      u^{-}_{\beta}({\bf j},t)u^{-}_{\gamma}({\bf k}-{\bf j},t)
  \nonumber \\
  &+& 2u^{-}_{\beta}({\bf j},t)u^{+}_{\gamma}({\bf k}-{\bf j},t)
      + u^{+}_{\beta}({\bf j},t)u^{+}_{\gamma}({\bf k}-{\bf j},t)
      \Big],
\label{NSE:up}
\end{eqnarray}
where the superscripts $+$ and $-$ on $M_{\alpha\beta\gamma}({\bf k})$
have the same significance as for $u_{\alpha}({\bf k},t)$, and the
symmetrized inertial transfer operator
$M_{\alpha\beta\gamma}({\bf k})
=(2i)^{-1}[k_{\beta}D_{\alpha\gamma}({\bf k})
+k_{\gamma}D_{\alpha\beta}({\bf k})]$.

In principle, the RG approach involves two stages:
(i) Eliminate the high-$k$ modes, ${\bf u}^+$, which appear in equation
(\ref{NSE:um}) for $0<k<k_1$, by solving for the mean
effect of the high-$k$ field. This results in an increment to the
viscosity, i.e. $\nu_0\rightarrow\nu_1=\nu_0+\delta\nu_0$.
(ii) Rescale the basic variables, so that the Navier-Stokes equation for
$0<k<k_1$ looks like the original Navier-Stokes equation for $0<k<k_0$.

Although this procedure is appealingly simple and has a clear physical
interpretation, it has not proved easy to put into practice in the
turbulence problem. A typical approach is to
eliminate all the high-$k$ effects in equation (\ref{NSE:um}), by
substituting the solution of equation (\ref{NSE:up}), directly into the
${\bf u}^+$  modes in the ${\bf u}^-$ equation. However,
problems are then encountered because of the mode coupling between
${\bf u}^-$ and ${\bf u}^+$. Even if one succeeds in carrying out the
first part, the further problem of averaging out the high-$k$ modes
arises immediately, because ${\bf u}^-$ and ${\bf u}^+$ are not
statistically independent. This problem was avoided by Foster, Nelson
and Stephen (1977; hereafter referred to as FNS) 
in their pioneering study of stirred fluid motion, as
they restricted their attention to stirring forces which were
multivariate normal and excluded the effects of the turbulence cascade.
However, it has been shown that the use of a `filtered' average by FNS
to eliminate the ${\bf u}^-$ equation is really an uncontrolled
approximation (Eyink, 1994).

\subsection{Iterative-Averaging RG with Results}

\begin{figure}
\centerline{\psfig{figure=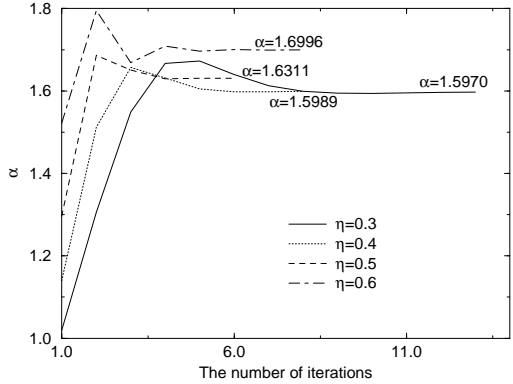,width=8truecm}}
\caption[]{\small\sf
         Convergence of the Kolmogorov spectral constant $\alpha$ to the
         fixed points for several values of the bandwidth parameter
         $\eta$.
        }
\label{alpha_vs_n}
\end{figure}

Here, we follow the method of iterative averaging, which is based upon
the derivation of a recurrence relation and, in principle, eliminating
finite blocks of modes (i.e. high-$k$ modes) while maintaining the form
invariance of the dynamical equation. Apart from the work of FNS,
elimination procedures can be performed by `conditional' averaging, first
introduced by McComb (1982). 
Further details about the conditional average have been given elsewhere
(McComb, Robert and Watt, 1992).
The basic ansatz of a conditional average is
that a small uncertainty ($\Phi^{-}$, say) at the cutoff wavenumber
will generate chaotic behaviour for the high-$k$ modes. Although the
introduction of $\Phi^{-}$ has been accepted, mainly due to the chaotic
nature of the Navier-Stokes equations, it might be interesting to see
how `rapidly' chaotic behaviour develops from the given small
$\Phi^{-}$ by numerical simulation. This aspect is one of our
current tasks and the results will be reported in due course. 

\begin{figure}
\centerline{\psfig{figure=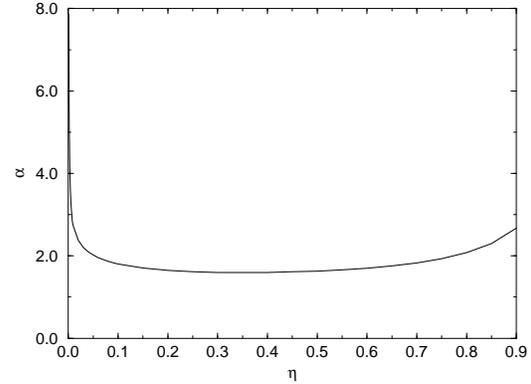,width=8truecm}}
\caption{\small\sf
         Dependence on the bandwidth parameter $\eta$ of the calculated
         values of the Kolmogorov spectral constant $\alpha$ based on
         equation (\ref{viscosity:final}).
        }
\label{alpha_vs_eta}
\end{figure}

The current result of the iterative-averaging calculation for the
Navier-Stokes equations after first eliminating the high-$k$ effects is
\begin{eqnarray}
\lefteqn{\left[\frac{\partial}{\partial t} + \nu_1 k^2\right]
         u^{-}_{\alpha}({\bf k},t)
        }
  \nonumber \\
  &=& M^{-}_{\alpha\beta\gamma}({\bf k})\int d^3 j~
      u^{-}_{\beta}({\bf j},t)u^{-}_{\gamma}({\bf k}-{\bf j},t),
\end{eqnarray}
where $\nu_1 = \nu_0 + \delta\nu_0(k)$ and
\begin{eqnarray}
& \delta\nu_0(k) &
  = -\frac{1}{k^2}\int d^3 j~
      Q^+_v({\bf j})
  \nonumber\\
  &\times&
\frac{\frac{4}{d-1}
      \mbox{{\bf Tr}}
      \big[
      M^{-}_{\alpha\beta\gamma}({\bf k})
      M^{+}_{\gamma\rho\sigma}({\bf k}-{\bf j})
      D_{\beta\sigma}({\bf j})
      \big]
     }
     {\nu_0 j^2 + \nu_0 |{\bf k}-{\bf j}|^2
     }.
\label{viscosity:final}
\end{eqnarray}
Here, we consider space dimension $d=3$. This result can be extended to
further shells, and we have shown elsewhere (McComb and Watt, 1992) that
a fixed point is reached under numerical iteration of the recursion
relations (see also Figure \ref{alpha_vs_n}). In Figure
\ref{alpha_vs_eta}, we show a calculation of the
Kolmogorov constant $\alpha=1.60\pm 0.01$ independent of the bandwidth
of modes being eliminated for bandwidths in the range 
$0.25\leq\eta\leq 0.45$, in agreement with  experiment.

\section{NUMERICAL SIMULATIONS}

Two programmes of numerical simulation are being carried out --- one at
the University of Edinburgh in the United Kingdom, the other at the
Swiss Federal Institute of Technology, Lausanne.  A large number of
runs have already been carried out at Lausanne, and this paper presents 
some of the results obtained so far.

The simulations themselves are very similar, while the computer systems 
on which they are run differ greatly.  At Edinburgh, work is carried out 
on a parallel machine, the Cray T3D, while in Lausanne a parallel-vector
machine, the NEC SX-4, is used.

The simulations discussed in this paper were carried out at a
resolution of $256^3$, requiring approximately 14 seconds
of SX-4 time per time-step on a single processor.

The general method of such simulations has been well established.
We follow the work of Orszag for the construction of initial velocity
fields (1969) and in the use of a pseudospectral method (1971).
The time integration scheme is a second-order Runge-Kutta method
and 
partial dealiasing is achieved by way of a random-shifting method
(see, for example, Rogallo, 1981).

\subsection{Initial Conditions}

The simulations are started with an initial energy spectrum of the form
\begin{equation}
E(k,0) = 16(2/\pi)^{\frac{1}{2}}u_0^2k_{\mbox{p}}^{-5}k^4
\exp[-2(k/k_{\mbox{p}})^2]
\end{equation}
where $k_{\mbox{p}}$ is the location of the spectrum's maximum and
$u_0$ is the required initial r.m.s. velocity.

\subsection{Forcing}

Stationary turbulence is obtained by use of a deterministic
forcing term
\begin{equation}
f_{\alpha} ({\bf k}, t) = \left\{ \begin{array}{l l}
\varepsilon u_\alpha({\bf k}, t) / (2E_f(t)) & \mbox{if $0<k<k_f$,} \\
0 & \mbox{otherwise,}
\end{array} \right. \label{equation for f_alpha}
\end{equation}
where $\varepsilon$ is the mean dissipation rate, and
\begin{equation}
E_f(t) = \int_0^{k_f} E(k,t) \mbox{d}k.
\end{equation}
There is no preferred direction in this forcing
and the turbulence rapidly reaches a statistically isotropic
and steady state.

\subsection{Statistics}

While our simulations are entirely conventional, we do not rely solely
on the usual practice (as justified by isotropy) of averaging
over shells in wavenumber space in order to obtain statistical
quantities, but also generate many realizations in order to
increase our sample size.

The main characteristics of the simulation
are reported in Table \ref{parans1},
where $\Delta t$ is the
time step, $T$ is the integration time, $\nu_o$ is the molecular
viscosity, $k_f$ is defined in (\ref{equation for f_alpha}),
$k_0$ is the ultraviolet cut-off, $\varepsilon$
is the mean dissipation rate, $R_\lambda$ is the Reynolds number
based on the Taylor microscale, $L$ is the integral scale,
$\lambda$ is the Taylor microscale, $\tau_E$ is the turnover time
and $s_3$ and $s_4$ are respectively the skewness and flatness
of the velocity derivative.
\begin{table}[t]
\begin{center}
\begin{tabular}{|c|c|c|c|c|c|}
\hline
$\Delta t$&$T$&$\nu_o$&$k_f$&$k_0$\\
\hline\hline
$10^{-3}$&$113.5$&$10^{-3}$&
$1.5$&$120$\\ \hline
\end{tabular}
\\
\vspace{5mm}
\begin{tabular}{|c|c|c|c|c|c|c|c|}
\hline
$\varepsilon$&$R_\lambda$&$L$&$\lambda$&$\tau_E$&$s_3$&$s_4$\\
\hline\hline
$.15$&$190.606$&$1.431$&$.246$&$1.853$&$-.51$&
$6.053$\\ \hline
\end{tabular}
\end{center}
\caption{\small\sf Characteristics of the simulation}
\label{parans1}
\end{table}

The equations have been integrated for more
than $60$ turnover times and about $200$ box-realizations
of each component of the velocity field have been stored in a database.
Since these box-realizations are separated by $\approx \tau_E/4$ they
can be considered statistically independent for the middle-range-scales
and the small-scales.

\section{RESULTS}

\begin{figure}
\centerline{\psfig{figure=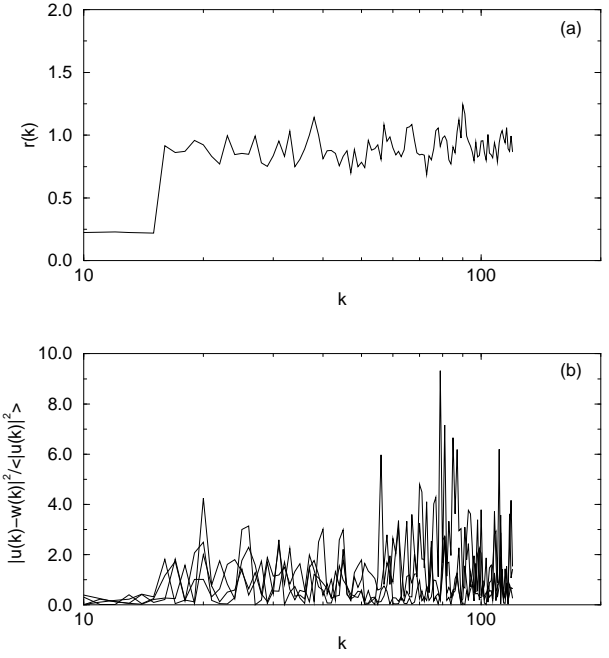,width=8truecm}}
\caption{\small\sf
         (a) Relative energy error for
             $k_b=10$, $k_c=15$, $\zeta=0.5$ and $\alpha=1$.
         (b) A selected set of realizations showing strong
             fluctuations for $k \geq 15$.
        }
\label{lucfig1}
\end{figure}

We wish to assess the freedom to carry out conditional
averages of the type required by RG.  In principle we
may do this by extracting,
from an ensemble of
realizations of the velocity field
\begin{eqnarray}
{\cal X}=\{X_{\alpha}^{(n)}(\Vec{k},t)\,|\,\alpha=1,2,3; t\in [0,T];
\nonumber \\
0\leq |\Vec{k}| \leq k_0; n=1,...,N\},
\end{eqnarray}
two disjoint
subensembles ${\cal Y}$ and ${\cal Z}$
chosen such that, for a prescribed $\zeta>0$,
\begin{eqnarray}
\frac{
|\Vec{Y}^{(m)}(\Vec{k},t)-\Vec{Z}^{(m)}(\Vec{k},t)|^2}
{2\Avr{|\Vec{Y}^{(m)}(\Vec{k},t)|^2}}\leq\zeta
\hspace{3mm} \nonumber \\
\mbox{for all}\hspace{3mm} 0\leq|\Vec{k}|\leq k_c\,;\,m=1,...,M\,;\,
t\in [0,T],
\end{eqnarray}
for each realization $\Vec{Y}^{(m)}\in {\cal Y}$ and
$\Vec{Z}^{(m)}\in {\cal Z}$.
We may then define
the relative energy of the error
\begin{equation}
r(|\Vec{k}|)=\frac{\Avr{(\Vec{u}(\Vec{k},t)-\Vec{w}(\Vec{k},t))^2}}
{2\Avr{\Vec{u}(\Vec{k},t)^2}}, \label{rk}
\end{equation}
where $\Vec{u}(\Vec{k},t)\in{\cal Y}$ and $\Vec{w}(\Vec{k},t)\in{\cal Z}$.
(It is important to note that the averages in the definition (\ref{rk})
are, in this context, subensemble averages defined on
${\cal Y}$ and ${\cal Z}$ and not ensemble averages on ${\cal X}$.)
In equation (\ref{rk}) and subsequently, we assume that the fields are
statistically stationary and isotropic, therefore $r$ depends only
on $|\Vec{k}|$.
Since the two fields are very close when $0\leq |\Vec{k}|\leq k_c$,
$r(|\Vec{k}|)$ will be much 
smaller than $1$ in this interval, indicating that
the fields are almost completely correlated.
If the error between the fields grows
in such a way that they become
decorrelated, we will have $r(|\Vec{k}|)\rightarrow 1$ as
$|\Vec{k}|\geq k_c$ increases.

In practice, our 200 box-realizations are not sufficient for
the above analysis and we shall describe how we have extracted, using a
partial sampling technique, enough realizations to compute the relative 
energy of the error defined by (\ref{rk}).

In order to this, we have performed the following partial
Fourier transform of one component of the velocity field
\begin{equation}
u_\alpha(x,y,k)=\frac{1}{2\pi}\int u_\alpha(x,y,z) e^{ikz}\,dk,
\end{equation}
then we have selected, for each box-realization, a set of realizations,
say
$u_\alpha(x_i,y_i,k)$, where the spacing
$\delta x=|x_{i+1}-x_i|=|y_{i+1}-y_i|$ is chosen such that the realizations
are (approximately) independent for the range of $k$ we consider (if we consider
only the scales such that $k\geq k_b$, then $\delta x=2\pi/k_b$). The union
of all these realizations obtained for each of the box-realizations will
constitute our ensemble ${\cal X}$. The subensemble ${\cal Y}$
is formed by choosing an arbitrary subensemble of ${\cal X}$. To select the
subensemble ${\cal Z}$, we impose the condition
\begin{eqnarray}
\frac{
|Y^{(m)}(k)-Z^{(m)}(k)|^2}
{2\Avr{|Y^{(m)}(k,t)|^2}}\leq\zeta
\hspace{3mm} \nonumber \\ 
\mbox{for all}\hspace{3mm} k_b\leq k\leq k_c\,;\,m=1,...,M.
\end{eqnarray}
Note that the time dependence does not appear in the equations
since
all the box-realizations used to form the
ensemble ${\cal X}$ are taken in the statistically steady
regime. Figure \ref{lucfig1}(a) shows the
relative energy error 
\begin{equation}
r(k)=\frac{\Avr{(u(k)-w(k))^2}}
{2\Avr{u(k)^2}},
\end{equation}
where $u\in {\cal Y}$ and $w\in {\cal Z}$ for 
$k_b=10$, $k_c=15$, $\zeta=0.5$ and $\alpha=1$.
The number of realizations $M$ is 2533. Though the number of realizations
is not large enough to have a smooth converged solution, one can see that
the relaxation to a chaotic regime is indeed very fast.
Figure \ref{lucfig1}(b)
shows a selected set of realizations for which one can observe that
the constraint imposed for $10\leq k\leq15$ does not prevent strong
fluctuations for $k\geq 15$. The convergence
of $r(k)$ is difficult to improve, due to the restriction on the number
of realizations available for a given constraint.

\begin{figure}
\centerline{\psfig{figure=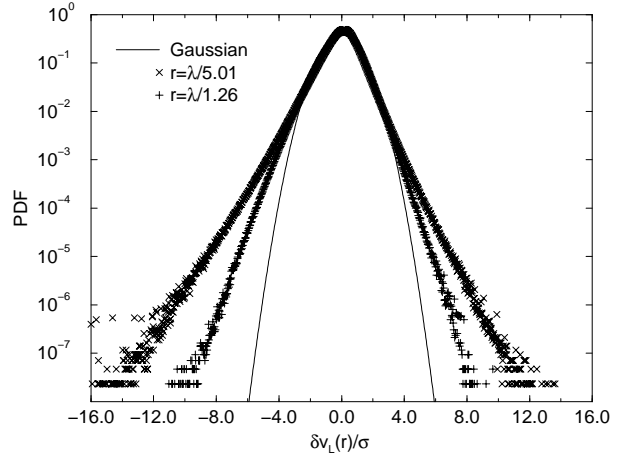,width=8truecm}}
\caption{\small\sf
         Normalized pdf of the unconditional ensemble.
        }
\label{lucfig2}
\end{figure}

\begin{figure}
\centerline{\psfig{figure=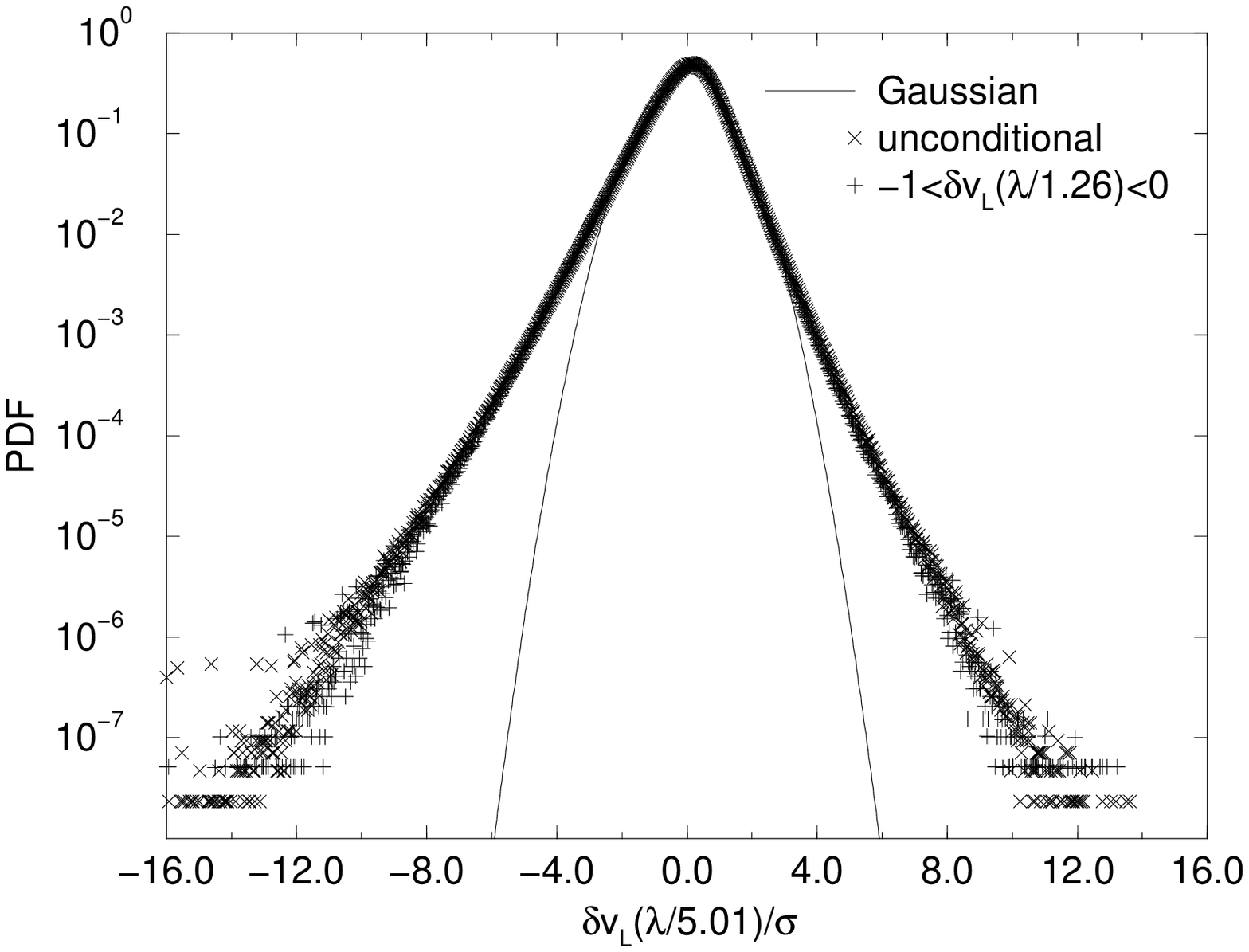,width=8truecm}}
\caption{\small\sf
         The pdfs of the unconditional ensemble and a subensemble
         defined by the constraint $-1<\delta v_L(h_1)<0$.
        }
\label{lucfig3}
\end{figure}

\begin{figure}
\centerline{\psfig{figure=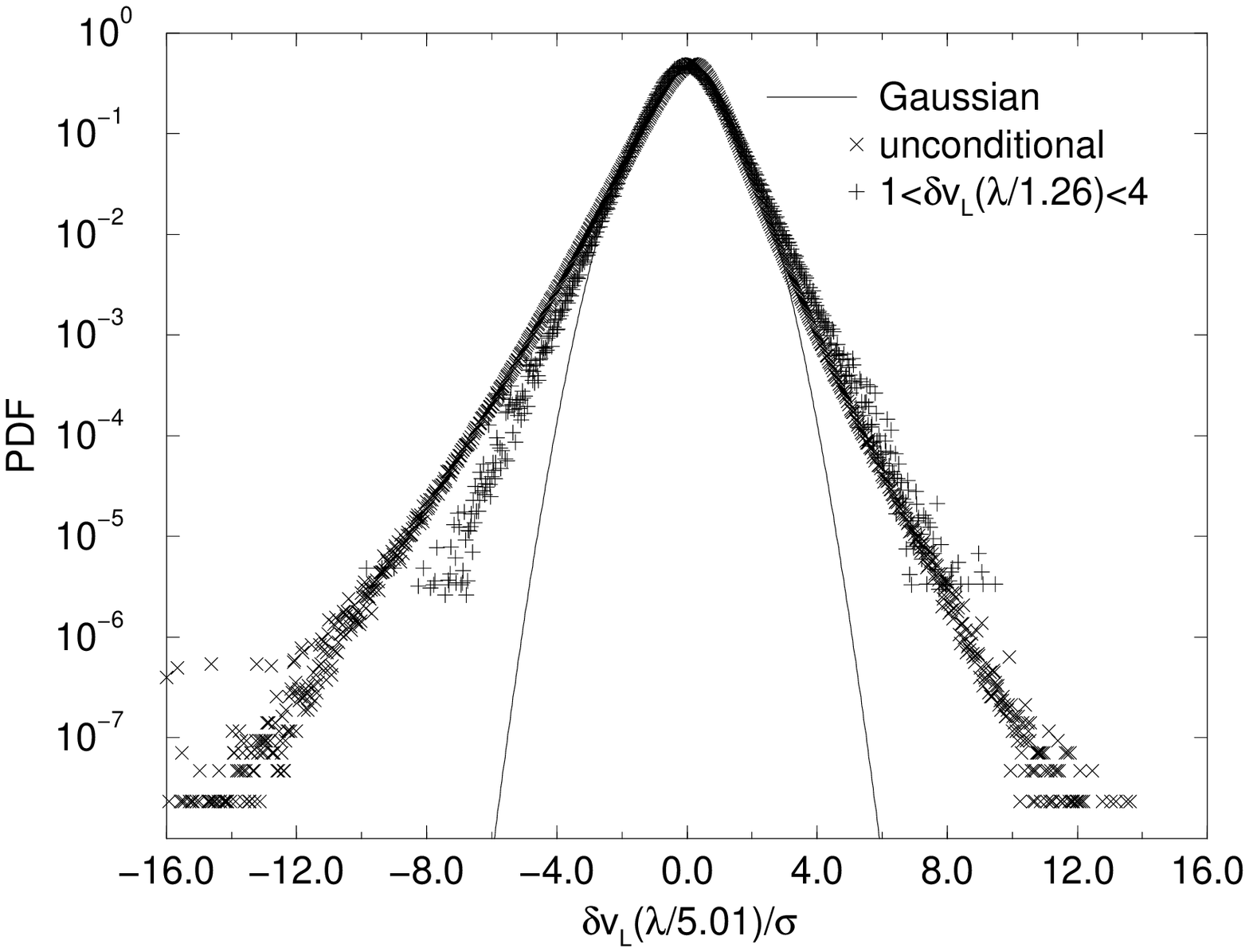,width=8truecm}}
\caption{\small\sf
         The pdfs of the unconditional ensemble and a subensemble
         defined by the constraint $1<\delta v_L(h_1)<4$.
        }
\label{lucfig4}
\end{figure}

\begin{figure}
\centerline{\psfig{figure=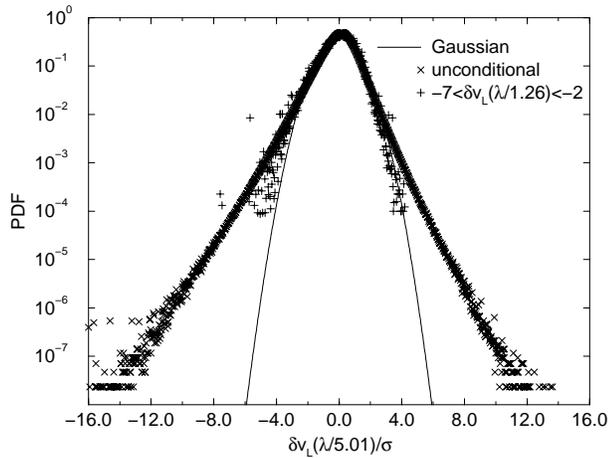,width=8truecm}}
\caption{\small\sf
         The pdfs of the unconditional ensemble and a subensemble
         defined by the constraint $-7<\delta v_L(h_1)<-2$.
        }
\label{lucfig5}
\end{figure}

Another natural way in which the
small-scale properties of a conditional subensemble may be
investigated is by studying the probability density
functions (pdfs) of velocity increments.  In physical-space, we
can use homogeneity in the three dimensions and have sufficiently
large subensembles to compute high-order statistics and pdfs.
The velocity increments are defined by the following
relation
\begin{equation}
\delta\Vec{u}(\Vec{x},\Vec{h})=\Vec{u}(\Vec{x}+\Vec{h})-\Vec{u}(\Vec{x}),
\end{equation}
where $\Vec{h}$ is a displacement vector and $\Vec{x}$ the position.
Since the fields are statistically isotropic, we can restrict ourselves
to the study of the longitudinal velocity increment
$\delta v_L(h)$ which is the projection of $\delta\Vec{u}(\Vec{h})$
on the direction of the vector $\Vec{h}$ and the lateral velocity
increment $\delta v_T(h)$ which is the projection of $\delta\Vec{u}(\Vec{h})$
on a direction perpendicular to $\Vec{h}$. For the purpose
of this paper, we have only studied
the longitudinal velocity increment $\delta v_L(h)$.
We have selected
two scales, $h_1=\lambda/1.26$ and $h_2=\lambda/5.01$
($\lambda$ is the Taylor micro-scale, therefore $h_1$ is a typical
scale in the
inertial subrange and $h_2$ is in the dissipation subrange). The
selection of the subensembles is performed using conditions
of the type $a< \delta v_L(h_1)< b$.  The pdfs of
$\delta v_L(h_2)$ for the unconditional ensemble
and for the subensembles are then compared.
Figure \ref{lucfig2} gives the normalized pdf ($\sigma$ is
the standard deviation of $\delta v_L(h)$) of the unconditional ensemble
for $h=h_1$ and $h=h_2$. We observe the classical result that
the tails of the pdfs are growing as the scale
is decreased which is the signature of growing intermittency.
The pdf also shows a negative skewness which is a direct consequence
of the nonlinear dynamics of the Navier-Stokes equations.
Figure \ref{lucfig3}, shows the pdfs of the unconditional ensemble and of
a subensemble defined by the constraint $-1<\delta v_L(h_1)<0$. The
pdfs are almost superimposed, showing that the flow at scale $h_2$
is unaffected by the condition imposed at scale $h_1$.
Figure \ref{lucfig4} is a case for which
the subensemble is much smaller due to a more restrictive
condition, $1<\delta v_L(h_1)<4$. However, the general behavior of
the pdf supports the view that the chaotic dynamics of the Navier-Stokes
equations tends to restore the original distribution. Note
that the skewness is incorrectly predicted and seems to be correlated with
the sign of $\delta v_L(h_1)$. Figure \ref{lucfig5} presents a case with a
very strong condition, $-7<\delta v_L(h_1)<-2$. Though the number
of realizations is small, we observe that the top of the pdf
is quite accurately reproduced.
 
\section{CONCLUSION}

These results, although preliminary in nature, offer crucial
support to the hypothesis that a conditional average may be
used to reduce the number of degrees of freedom required
for the numerical simulation of turbulence.  Work is continuing
to make a more stringent assessment of the validity of such 
averages for turbulence and this includes carrying out simulations
at higher numerical resolution.  At present we are working on
a $512^3$ simulation and hope to present results from this at the
conference.

\vspace{0.5truecm}
\noindent{\sl Acknowledgements:} 
The simulation presented in this paper has been
performed on the computers of the Swiss Center
for Scientific Computing, Manno. The research of
L. Machiels is supported by the Swiss National
Foundation for Scientific Research.

\section{REFERENCES}

Eyink, G. L., 1994, ``Renormalization group method in statistical
hydrodynamics'', {\sl Phys. Fluids}, Vol. 6, pp. 3063--3078.

Forster, D., Nelson, D. R. and Stephen, M. J., 1977, ``Large-distance
and long-time properties of a randomly stirred fluid'',
{\sl Phys. Rev. A}, Vol. 16, pp. 732--749.

McComb, W. D., 1982, ``Reformulation of the statistical equations for
turbulent shear flow'', {\sl Phys. Rev. A}, Vol. 26, pp. 1078--1094.

McComb, W. D. 1990, {\sl The Physics of Fluid Turbulence}, Oxford University
Press.

McComb, W. D. 1995, ``Theory of Turbulence'', {\sl Rep. Prog. Phys.},
Vol. 58, No. 10, pp. 1117--1205.

McComb, W. D., Robert, W. and Watt, A. G., 1992, ``Conditional-averaging
procedure for problems with mode-mode coupling'', {\sl Phys. Rev. A},
Vol. 45, pp. 3507--3515.

McComb, W. D. and Shanmugasundaram, V. 1983. "Some developments in the
application of renormalization methods to turbulence theory". Paper
presented to the {\em Fourth Symp. on Turb. Shear Flows}, Karlsruhe,
Germany September 12--14, 1983.

McComb, W. D., "Some recent developments in the application of
renormalization methods to problems in turbulence." Paper presented to
the {\em Eighth Symp. on Turb. Shear Flows}, Munich, Germany, September
9--11, 1991.

McComb, W. D. and Watt, A.G., 1992, ``Two-field theory of
incompressible-fluid turbulence'', {\sl Phys. Rev. A}, Vol. 46,
pp. 4797--4812.

Orszag, S., 1969, ``Numerical Methods for the Simulation of Turbulence,''
{\sl Phys. Fluids (suppl. 2)}, Vol. 12, pp. 250--257.

Orszag, S., 1971, ``Numerical Simulation of Incompressible Flows Within
Simple Boundaries.  I. Galerkin (Spectral) Representations,''
{\sl Stud. Appl. Maths.}, Vol. 50, No. 4, pp. 293--327.

Rogallo, R.S., 1981, ``Numerical Experiments in Homogeneous Turbulence,''
NASA TM-81315.

\end{document}